# Improving Generative Adversarial Networks for Video Super-Resolution

**Daniel Wen**
dywen@ucsc.edu

## Abstract

In this research, we explore different ways to improve generative adversarial networks for video super-resolution tasks from a base single image super-resolution GAN model. Our primary objective is to identify potential techniques that enhance these models and to analyze which of these techniques yield the most significant improvements. We evaluate our results using Peak Signal-to-Noise Ratio (PSNR) and Structural Similarity Index (SSIM). Our findings indicate that the most effective techniques include temporal smoothing, long short-term memory (LSTM) layers, and a temporal loss function. The integration of these methods results in an 11.97% improvement in PSNR and an 8% improvement in SSIM compared to the baseline video super-resolution generative adversarial network (GAN) model. This substantial improvement suggests potential further applications to enhance current state-of-the-art models.

## 1 Introduction

As machine learning and artificial intelligence engulfs the computer engineering field in recent years, one popular and useful computer vision task is image super-resolution; the goal is to increase the resolution of an image, often by a factor of 4x or more, while maintaining its content details as much as possible. Similarly, video super-resolution is an extension of this where the task is to increase the resolution of a video sequence, typically from lower to higher resolutions. Real-world applications to video super-resolution include enhancing the resolution of images and videos for medical imaging, surveillance and security, video conferencing, autonomous vehicles, entertainment and media, and many more.

In this paper, we aim to explore and gain proficiency with super-resolution GAN architectures by transforming a simple single-image super-resolution model into a video super-resolution model and achieving a significant performance improvement. Our primary objective is to develop expertise rather than to utilize current state-of-the-art models. However, the insights and techniques derived from our research could potentially be applied to enhance existing state-of-the-art models and contribute to the advancement of the field.

## 2 Related Work

### 2.1 Generative Adversarial Networks

Generative Adversarial Networks (GANs) represent a groundbreaking advancement in the field of machine learning and artificial intelligence, first introduced by Ian Goodfellow and his colleagues in 2014(3). GANs consist of two primary components: the generator and the discriminator. In the context of video super-resolution, the generator's role is to upscale low-resolution video frames to higher resolution, aiming to produce images that are perceptually indistinguishable from true high-resolution frames. The discriminator, on the other hand, evaluates both the real high-resolution frames and the generated frames, attempting to discern between the two. Through this adversarial process, the generator is continually refined to produce more realistic and detailed frames. This framework is particularly advantageous for video super-resolution because it encourages the generation of high-frequency details and textures, which are crucial for achieving visually appealing results. Additionally, advanced variants of GANs, such as recurrent GANs or those incorporating temporal coherence mechanisms, further enhance the model's ability to maintain consistency across video frames, addressing challenges unique to video data. This adversarial approach, therefore, provides a robust and effective solution for enhancing video resolution. In addition to the architecture, the paper also discusses the advantages and disadvantages of using GANs compared to previous modeling frameworks. Additionally, it elaborates on the theoretical outcomes expected from the mini-batch stochastic gradient descent training of GANs and presents experimental results on the MNIST, TFD, and CIFAR-10 datasets. The innovative framework of GANs has since catalyzed significant progress in various applications, including image synthesis, video generation, and super-resolution, making it a pivotal technique in contemporary deep learning research. This architecture serves as the foundation for our model's approach to video super-resolution.

### 2.2 Image Super-Resolution

While there are several studies exploring various approaches to super-resolution, such as Kappeler et al.(5), one of the most promising results is derived from Ledig et al.'s SRGAN approach(4). In their work, they introduce a single-image super-resolution model utilizing residual blocks comprising convolutions, batch normalizations, PReLUs, and elementwise summation for the generator network. The discriminator network is constructed using simpler blocks consisting of single convolutions, batch normalizations, and leaky ReLU activations. By applying this model to single-image inputs, Ledig et al. established a new state-of-the-art in image super-resolution. We adopt this approach as the foundation for our video super-resolution model.

## 3 Approach

### 3.1 Modifying Generator Architecture

```python
# Define motion estimation function
def estimate_motion_vectors(frames):
    motion_vectors = []
    prev_gray = cv2.cvtColor(frames[0], cv2.COLOR_BGR2GRAY)

    for i in range(1, len(frames)):
        gray = cv2.cvtColor(frames[i], cv2.COLOR_BGR2GRAY)
        flow = cv2.calcOpticalFlowFarneback(prev_gray, gray, None,
            0.5, 3, 15, 3, 5, 1.2, 0)
        motion_vectors.append(flow)
        prev_gray = gray

    return motion_vectors

# Define motion vector smoothing function
def smooth_motion_vectors(motion_vectors, alpha=0.9):
    smoothed_vectors = [motion_vectors[0]]

    for i in range(1, len(motion_vectors)):
        smoothed_vector = alpha * motion_vectors[i] + (1 - alpha) *
            smoothed_vectors[-1]
        smoothed_vectors.append(smoothed_vector)

    return smoothed_vectors

# Define frame alignment function
def align_frames(frames, smoothed_vectors):
    aligned_frames = [frames[0]]

    for i in range(1, len(frames)):
```

```
29            flow = smoothed_vectors[i-1]
30            h, w = flow.shape[:2]
31            flow_map = np.column_stack((np.repeat(np.arange(h), w), np.
                tile(np.arange(w), h))) + flow.reshape(-1, 2)
32            remap_frame = cv2.remap(frames[i], flow_map[:, 0].reshape(h, w
                ).astype(np.float32),
33                                    flow_map[:, 1].reshape(h, w).astype(np
                                        .float32), cv2.INTER_LINEAR)
34            aligned_frames.append(remap_frame)
35
36        return aligned_frames
```

Listing 1: Motion Estimation, Smoothing, and Frame Alignment Functions

In this study, we utilize a visual modality input to our model analogous to our base single image super-resolution model, but with frames extracted from the input video fed as a sequence of images. To maintain temporal consistency across the sequence of frames, we integrate a long short-term memory (LSTM) layer in the generator network2. Unlike single image super-resolution tasks, where shuffling input frames may be beneficial, we retain the sequence order in all training and test datasets to ensure the model effectively captures the video's temporal features. Additionally, we incorporate temporal smoothing techniques?? by maintaining motion continuity between frames, employing Gaussian smoothing to the motion vectors to average them with a Gaussian-weighted sum. Finally, a temporal loss function is added to ensure coherence between consecutive frames.

```
1  class Generator(nn.Module):
2      def __init__(self, scale_factor, sequence_length):
3          upsample_block_num = int(math.log(scale_factor, 2))
4
5          super(Generator, self).__init__()
6          self.sequence_length = sequence_length
7
8          self.conv1 = nn.Sequential(
9              nn.Conv2d(3, 64, kernel_size=9, padding=4),
10             nn.PReLU()
11         )
12         self.res_blocks = nn.Sequential(
13             *[ResidualBlock(64) for _ in range(sequence_length)]
14         )
15         self.conv2 = nn.Sequential(
16             nn.Conv2d(64, 64, kernel_size=3, padding=1),
17             nn.BatchNorm2d(64)
18         )
19         self.upsample_blocks = nn.Sequential(
20             *[UpsampleBLock(64, 2) for _ in range(upsample_block_num)]
21         )
22         self.conv3 = nn.Conv2d(64, 3, kernel_size=9, padding=4)
23
24         # LSTM layer for temporal consistency
25         self.lstm = nn.LSTM(input_size=64 * 64 * 64, hidden_size=256,
                num_layers=1, batch_first=True)
26         self.fc = nn.Linear(256, 64 * 64 * 64)
27
28     def forward(self, x):
29         batch_size, seq_len, c, h, w = x.size()
30         x = x.view(batch_size * seq_len, c, h, w)
31
32         conv1 = self.conv1(x)
33         res_blocks = self.res_blocks(conv1)
34         conv2 = self.conv2(res_blocks)
35         res_out = conv1 + conv2
36
37         res_out = res_out.view(batch_size, seq_len, -1)
38         lstm_out, _ = self.lstm(res_out)
```

```
39              lstm_out = lstm_out.contiguous().view(batch_size * seq_len,
                  -1)
40              fc_out = self.fc(lstm_out)
41              fc_out = fc_out.view(batch_size * seq_len, 64, 64, 64)
42  
43              upsampled = self.upsample_blocks(fc_out)
44              output = self.conv3(upsampled)
45              return (torch.tanh(output) + 1) / 2
```

Listing 2: Generator Network for Video Super-Resolution

### 3.2 Generator Loss Function

The generator's loss function3 consists of adversarial loss, perceptual loss, image loss, total variation (TV) loss, and temporal consistency loss. The adversarial loss evaluates the generator's ability to deceive the discriminator, thereby encouraging the production of more realistic images. The perceptual loss employs a pre-trained VGG16 network to extract high-level feature representations from the images, highlighting the importance of features most noticeable to human perception. The image loss, measured as the mean squared error (MSE) between the generated and ground truth images, ensures that the generated images closely match the ground truth in terms of pixel values. The TV loss promotes spatial smoothness in the generated images by penalizing large differences between neighboring pixels, thus reducing noise. Finally, the temporal consistency loss ensures coherence between consecutive frames by penalizing discrepancies between super-resolved frames and preceding frames.

```
1   class GeneratorLoss(nn.Module):
2       def __init__(self):
3           super(GeneratorLoss, self).__init__()
4           vgg = vgg16(pretrained=True)
5           loss_network = nn.Sequential(*list(vgg.features)[:31]).eval()
6           for param in loss_network.parameters():
7               param.requires_grad = False
8           self.loss_network = loss_network
9           self.mse_loss = nn.MSELoss()
10          self.tv_loss = TVLoss()
11          self.temporal_loss = nn.MSELoss()  # Use MSE loss for temporal
                consistency
12  
13      def forward(self, out_labels, out_images, target_images,
            out_images_prev=None, target_images_prev=None):
14          # Adversarial Loss
15          adversarial_loss = torch.mean(1 - out_labels)
16  
17          # Perceptual Loss
18          perception_loss = self.mse_loss(self.loss_network(out_images),
                self.loss_network(target_images))
19  
20          # Image Loss
21          image_loss = self.mse_loss(out_images, target_images)
22  
23          # TV Loss
24          tv_loss = self.tv_loss(out_images)
25  
26          # Temporal Consistency Loss
27          if out_images_prev is not None and target_images_prev is not
                None:
28              temporal_consistency_loss = self.temporal_loss(out_images
                    - out_images_prev, target_images - target_images_prev)
29          else:
30              temporal_consistency_loss = 0.0
31
```

```
32          return image_loss + 0.001 * adversarial_loss + 0.006 *
                perception_loss + 2e-8 * tv_loss + 0.1 *
                temporal_consistency_loss
```

Listing 3: Generator Loss Function for Video Super-Resolution

## 4 Results

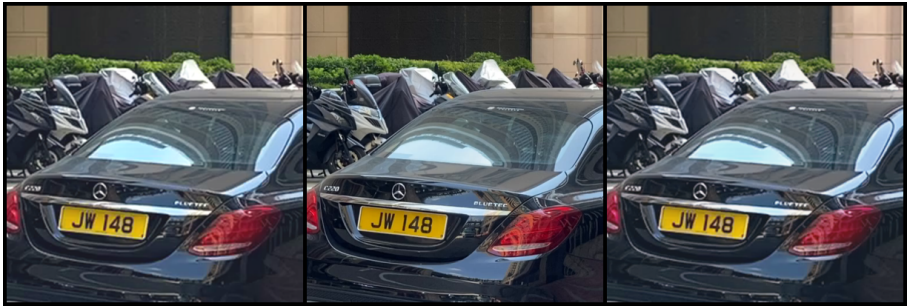

Figure 1: Comparison of Images. (Left) Ground Truth, (Middle) High-Resolution Image Generated by Our Model, (Right) Low-Resolution Counterpart.

| Model | PSNR | SSIM |
| --- | --- | --- |
| Base Video SRGAN | 16.32 | 0.41 |
| Base Video SRGAN + LSTM | 22.89 | 0.75 |
| Improved Video SRGAN | 25.63 | 0.81 |

Figure 2: SRGAN results

For training, we utilized the paired videos in the "videos.zip" file from the RealVSR dataset (1). The low-quality videos are used as inputs, and their paired high-quality videos serve as the desired outputs. For testing, we use the low-quality sequences found in "LQ_test.zip" and evaluate the performance using the corresponding ground-truth data. Each training lasts for 50 epochs with a learning rate of 0.0001, and an upscale value of 4x.

For evaluation, we used the Peak Signal-to-Noise Ratio (PSNR) and the Structural Similarity Index (SSIM). PSNR is a widely used metric for assessing the quality of reconstructed images and videos by comparing the similarity between the original and the reconstructed images. It is measured in decibels (dB) and quantifies the ratio between the maximum possible power of a signal and the power of the corrupting noise that affects the fidelity of its representation. Higher PSNR values typically indicate better quality, as they signify that the reconstructed image is closer to the original. On the other hand, SSIM is designed to measure the perceived quality of images by taking into account changes in structural information, luminance, and contrast. It compares local patterns of pixel intensities that have been normalized for luminance and contrast, providing a more comprehensive assessment of image quality. SSIM values range from 0 to 1, where a value closer to 1 indicates a higher similarity to the original image. Together, PSNR and SSIM offer a robust evaluation framework for assessing the effectiveness of super-resolution algorithms like SRGAN, capturing both pixel-level accuracy and perceptual quality.

Initially, with the transition from single image super-resolution to video super-resolution, the base model outputs a PSNR of 16.32 and an SSIM of 0.41. These suboptimal results are expected, as the base model is not optimized for video inputs. Upon incorporating the LSTM layer in the generator to maintain temporal consistency across the sequence of frames, we observe a PSNR of 22.89 and an SSIM of 0.75. While these results indicate a notable improvement and are more in line with expectations for a video super-resolution model, they still fall short of current state-of-the-art models.

Finally, by applying all our additional techniques, we achieve a PSNR of 25.63 and an SSIM of 0.81, representing an 11.97% improvement in PSNR and an 8% improvement in SSIM compared to the baseline video super-resolution generative adversarial network (GAN) model.

## 5 Conclusion

Our findings demonstrate successful attainment of our objective to explore video super-resolution GAN architectures. However, certain concerns warrant attention before applying our insights to state-of-the-art models. One particular concern is the behavior of the discriminator's loss function. Ideally, the discriminator's loss should converge around 0.5, indicating its ability to effectively distinguish between real (0.0) and generated (1.0) images. Concurrently, the generator's loss should tend towards 0, signifying its capability to generate realistic images. While our model achieves satisfactory convergence of the generator's loss, the discriminator's loss exhibits an upward trend, reaching 1.0 after 50 epochs. This suggests that the generator is producing images that consistently deceive the discriminator. To address this issue, enhancing the discriminator's architecture is proposed, potentially by augmenting the number of filters in its convolutional layers or integrating a dropout layer post the main convolutional blocks. Resolving these challenges promises to not only improve the performance of our video super-resolution model but also advance current state-of-the-art methodologies in the field.